\documentclass[amsmath,amssymb,prl,hyperlink,twocolumn]{revtex4}

	\usepackage{graphicx}
	\usepackage{soul}
	\usepackage[colorlinks=true,citecolor=blue,linkcolor=magenta]{hyperref}
	\usepackage[usenames]{color}
	\usepackage{amsfonts}
	\usepackage{color}
	\usepackage{booktabs}
	\usepackage{multirow}
	\usepackage{float}
	
\usepackage{times}

\begin{document}

\title{Near Transform-Limited Single Photons from an Efficient Solid-State Quantum Emitter}

	\author{Hui Wang,$^{1,2,3}$ Z.-C. Duan,$^{1,2,3}$ Y.-H. Li,$^{1,2,3}$ Si Chen,$^{1,2,3}$ J.-P. Li,$^{1,2,3}$ Y.-M. He$^{1,4}$, M.-C. Chen,$^{1,2,3}$ Yu He,$^{1,2,3}$ X. Ding,$^{1,2,3}$, Cheng-Zhi Peng,$^{1,2,3}$ Christian Schneider$^4$, Martin Kamp$^4$, Sven H\"{o}fling$^{1,4,5}$, Chao-Yang Lu$^{1,2,3}$ and Jian-Wei Pan$^{1,2,3}$ \vspace{0.2cm}}
	
\affiliation{$^1$ Shanghai Branch, National Laboratory for Physical Sciences at Microscale and Department of Modern Physics, University of Science and Technology of China, Shanghai, 201315, China}
\affiliation{$^2$ CAS Center for Excellence and Synergetic Innovation Center in Quantum Information and Quantum Physics, University of Science and Technology of China, Hefei, Anhui 230026, China}
\affiliation{$^3$ CAS-Alibaba Quantum Computing Laboratory, Shanghai, 201315, China}
\affiliation{$^4$ Technische Physik, Physikalisches Instit\"{a}t and Wilhelm Conrad R\"{o}ntgen-Center for Complex Material Systems, Universitat W\"{u}rzburg, Am Hubland, D-97074 W\"{u}zburg, Germany}
\affiliation{$^5$ SUPA, School of Physics and Astronomy, University of St. Andrews, St. Andrews KY16 9SS, United Kingdom}

\date{\today \vspace{0.1cm}}
	
	\begin{abstract}
 By pulsed $s$-shell resonant excitation of a single quantum dot-micropillar system, we generate long streams of a thousand of near transform-limited single photons with high mutual indistinguishability. Hong-Ou-Mandel interference of two photons are measured as a function of their emission time separation varying from 13$\,$ns to 14.7$\,$${\mu}$s, where the visibility slightly drops from 95.9(2)\% to a plateau of 92.1(5)\% through a slow dephasing process occurring at time scale of 0.7$\,$$\mu$s. Temporal and spectral analysis reveal the pulsed resonance fluorescence single photons are close to transform limit, which are readily useful for multi-photon entanglement and interferometry experiments.
	\end{abstract}
	

	\maketitle

Self-assembled InGaAs quantum dots (QD) are promising single-photon emitters with high quantum efficiency and fast decay rate \cite{1.QDs}. In the past decades, extensive efforts have been devoted to producing single photons with high purity (that is, vanishing two-photon emission probability), near-unity indistinguishability, and high extraction efficiency \cite{2.SPS-1, 2.SPS-2, micropillar, pc, nanowire, two-photon, 2013NatComm, pulsedRF, ARP, 2016NatPho}. These key properties have been compatibly combined simultaneously on the same QD-micropillar very recently \cite{3.PerfectSPS, others}.

An important next challenge is to extend the single-photon sources to multiple photonic quantum bits \cite{4.RMP}, as required by various quantum information protocols such as boson sampling \cite{5.BoSam}, quantum teleportation \cite{xilin}, quantum computation \cite{6.OpticalQC}, and quantum metrology \cite{7.Metrology}. To this aim, one approach is to use many independent QDs \cite{8.manyQDs} which are tuned into identical emission wavelength \cite{8.QDtuning} and efficiently emit single photons stringently at the transform limit, that is, $T_2$$\,=\,$$2T_1$, where $T_2$ and $T_1$ are the photon's coherence time and lifetime, respectively. Another---probably less demanding---solution is based on only one \emph{perfect} QD emitting single-photon pulse trains with high efficiency \cite{3.PerfectSPS,others}, which are then either demultiplexed into $N$ spatial modes or dynamically controlled using time-bin encoding in a loop-based architecture \cite{9.timebin}. Implementing $N$-photon quantum circuits in this configuration demands streams of $N$ mutually indistinguishable single photons far apart in emission time.

\begin{figure}[tb]
    \centering
        \includegraphics[width=0.49\textwidth]{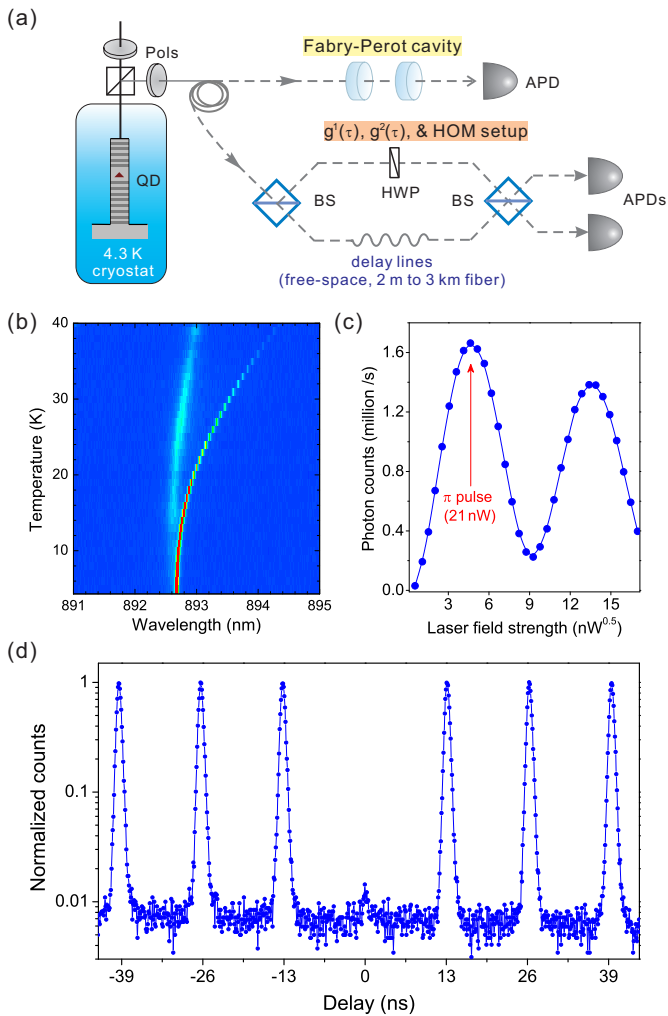}
\caption{Generation and characterization of single photons from a QD embedded in a micropillar. (a) An illustration of the experimental setup. The QD is grown by molecular beam epitaxy and sandwiched between 25.5 (15) lower (upper) distributed Bragg reflectors stacks. Pillars with a diameter of 2$\,$$\mu$m are defined via electron beam lithography \cite{device}. The quality factor of the micropillar cavity is measured to be 5349 (see Supplemental Fig.$\,$S1). The emitted RF single photons are sent to a Fabry-Perot cavity for high-resolution spectral analysis, or to an unbalanced Mach-Zehnder interferometer for $g^1(\tau)$, $g^2(\tau)$, and HOM measurements. (b) Temperature-dependent 2D intensity plot of the QD photoluminescence and the cavity mode. The excitation laser is at 780$\,$nm with a power of $\sim\,$0.3$\,$nW. (c) Detected single photon counts as a function of the pump field strength. (d) Intensity-correlation histogram of the RF photons. The measured second-order correlation at zero delay is $g^2(0)=0.007(1)$.}
\label{fig:1}
\end{figure}

However, previous Hong-Ou-Mandel (HOM) type interference experiments \cite{two-photon, pulsedRF, 3.PerfectSPS, others,2013NatComm,ARP,2016NatPho} were performed with time separation of only a few nanoseconds between two photons emitted consecutively from a QD. Spectral diffusion \cite{diffusion} with a time scale much slower than nanoseconds were speculated---yet without conclusive study---to account for the mismatch between the observed near-unity transient indistinguishability and the non-unity time-averaged $T_2/2T_1$ ratio \cite{two-photon, pulsedRF,2013NatComm,ARP,2016NatPho}. Thus it is highly desirable to study the two-photon interference as a function of their emission time separation and test how far apart can the high indistinguishability persists. The ultimate goal is to generate efficient and truly transform-limited single photons, with which perfect interference can be achieved regardless of their time separation, and even if the photons are from independent QDs.

An interesting remedy \cite{10.Mete} to circumvent the environment-induced dephasing is to operate in the small (typically $<10\%$) Rabi frequency regime of resonance fluorescence (RF) \cite{11.RF}. Under weak excitation, indistinguishable single photons with subnatural linewidth and phase locked to the excitation laser have been probabilistically generated \cite{10.Mete}, which found applications in solid-state quantum networking \cite{8.QDnetwork}. Another work revealed signatures of near transform-limited optical linewidth using resonant laser spectroscopy \cite{10.warburton}, yet without direct measurement of the emitted photons. Importantly, in these work \cite{10.Mete,10.warburton}, the  photon generation and collection efficiencies were intrinsically limited due to both the sample structures and the nondeterministic continuous-wave excitation. More recently, non-resonant pumping QDs grown by metal-organic chemical vapor deposition has generated single photon streams whose mutual indistinguishability decreases rapidly from 94(6)\% at 2$\,$ns separation to 53(3)\% at 12.5$\,$ns separation \cite{12.stephan}.

In this Letter, we demonstrate for the first time that a pulsed resonantly driven single QD embedded in a micropillar emits long streams of at least a thousand of single photons with high ($>\,$90\%) mutual indistinguishability, showing its promise for multi-photon experiments. Time-dependent HOM experiments reveal an indistinguishability of $95.9(2)\%$ for two single photons separated by 13$\,$ns, which decreases to a plateau of $92.1(5)\%$ at $\sim\,$2$-$14.7$\,$$\mu$s separation, through a dephasing process occurring at sub-$\mu$s time scale. Temporal and spectral measurements further confirm the single photons are close to transform limit: $T_2/2T_1$$\,=\,$$0.91(5)$.

Our experimental arrangement is shown in Fig.$\,$1a. A single InAs/GaAs self-assembled QD embedded inside a 2$\,$$\mu$m diameter micropillar cavity (for more details on sample growth and fabrication, see \cite{details}) is cooled to 4.3$\,$K where the QD emission is resonant with the cavity mode (see Fig.$\,$1b). The QD-micropillar is excited by a resonant, picosecond laser (see ref.$\,$\cite{3.PerfectSPS} for details), allowing a clear observation of Rabi oscillation in the detected RF photons as a function of laser field strength. At $\pi$ pulse with a repetition rate of 76.4 MHz, the QD-micropillar emits $\sim\,$5 million pulsed RF single photons at the output of a single-mode fiber (absolute source brightness $\sim\,$6.6\%), of which 1.67 million is finally detected by a silicon single-photon detector. The single-photon nature of the generated pulsed RF is unambiguously proven from intensity-correlation measurements \cite{HBT} (see Fig.$\,$2d) which show that at zero delay the multi-photon probability is almost vanishing ($g^2(0)=0.007(1)$).

The main result of this experiment is to measure the indistinguishability of the pulsed RF single photons as a function of their emission time separation ($\Delta$), and test how far apart can two photons still remain indistinguishable. As shown in Fig.$\,$1a, the single photons are fed into an unbalanced Mach-Zehnder interferometer with its path length difference variable from 2$\,$m to 3$\,$km optical fibers. We first test the HOM interference between two consecutively emitted single photons at a time delay of 13$\,$ns---the laser pulse separation. If two identical photons are combined on a beam splitter, they will always exit the beam splitter together through the same output port, a unique quantum phenomena that cannot explained by classical optics \cite{HOM}. Figure$\,$2a shows time-delayed histograms of normalized two-photon coincidence counts for cross (black) and parallel (red) polarization. A significant suppression of the counts is observed at zero delay when the two incoming photons are prepared in the parallel polarization state. We obtain a degree of indistinguishability of 0.959(2) between the two $\pi$-pulse excited single photons separated by 13$\,$ns. Reducing the laser excitation power can yield slightly higher indistinguishability up to $\sim\,$0.982(3) (see Supplemental Fig.$\,$S2), possibly due to a decrease of excitation induced dephasing \cite{EID}.

\begin{figure*}[tb]
    \centering
        \includegraphics[width=0.98\textwidth]{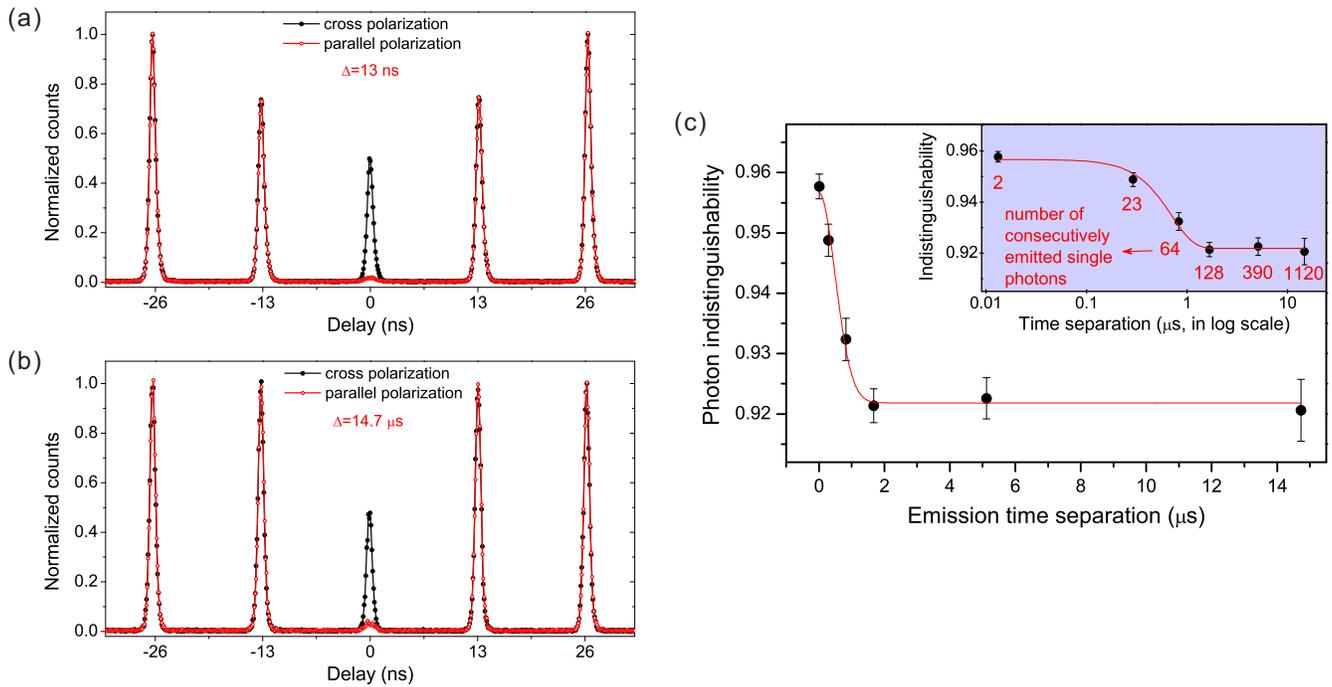}
\caption{Single photon indistinguishability measured from HOM experiments at various photon emission time separation ($\Delta$). (a) $\Delta$=13$\,$ns. The input two photons are $\pi$-pulse excited and prepared in cross (black) and parallel (red) polarizations, respectively. The data accumulation time was 5 minutes. The data points presented are raw data without background subtraction. (b) $\Delta$=14.7$\,$$\mu$s, while keeping all the other physical conditions the same as in (a). (c) The extracted photon indistinguishability as a function of emission time separation. The small window of few percent difference in the indistinguishability is precisely measured with a small error bar, which is possible due to the high photon count rate (that eliminates the shot noise) from the efficient QD-micropillar device \cite{3.PerfectSPS}. The same data is shown in the inset that plots the $x$ axis in log scale and highlights the number of consecutively emitted single photons for each data point. The red curve is a fit using the model derived in ref.$\,$\cite{12.stephan} assuming non-Markovian noise.}
\label{fig:2}
\end{figure*}

Similar experiments are performed by increasing the time separation to 289$\,$ns, 0.83$\,$$\mu$s, 1.67$\,$$\mu$s, 5.11$\,$$\mu$s, and 14.7$\,$$\mu$s, which is 3$-$4 orders of magnitude longer than the previous work \cite{two-photon, pulsedRF, 3.PerfectSPS, others,12.stephan}. The measurement result for the furthest separation of 14.7$\,$$\mu$s---corresponding to the HOM interference between the $n$-th and the $(n$$+$$1120)$-th photon in the pulsed train---is plotted in Fig.$\,$2b, which shows a photon indistinguishability of 0.921(5), only a small decrease from the figure of merit at $\Delta$=13$\,$ns. This finding demonstrates that the high mutual indistinguishability can be sustained among at least a thousand of single photons from a QD-micropillar, making it particularly suitable for optical quantum computing protocols with time-bin encoding \cite{9.timebin}.

Figure$\,$2c summaries the measured indistinguishability as a function of photon emission time separations, and the inset highlights the number of consecutively emitted single photons for each data point. The photon indistinguishability slightly drops from 95.9(2)\% at 13$\,$ns to a plateau of 92.1(5)\% at $\sim\,$2$-$14.7$\,$$\mu$s time separation. The time-dependent HOM measurements reveal a slow dephasing process occurring at a timescale of $\sim\,$0.7$\,$$\mu$s, which is $\sim\,$4000 times longer than the single-photon radiative decay lifetime. Such a dephasing process can be caused by spectral diffusion due to charge fluctuations in the vicinity of the QD \cite{diffusion}.

We expect that two single photons emitted with a time separation much shorter than 0.7$\,$$\mu$s collectively feel the same environment, thus their indistinguishability is immune to the slow spectral wandering and thus near-unity visibilities were observed in the HOM measurements \cite{pulsedRF, 3.PerfectSPS, others}. However, two photons separated much longer than microseconds experience different electric field at their time of emission, therefore their time averaged wave-packet overlap is determined by the amplitude of the spectral wandering compared to its intrinsic lifetime-limited linewidth.

Surprisingly, the photon indistinguishability maintains at a plateau of 92.1\% in the long time separation regime, suggesting that the emitted single photons can be close to transform limit. To confirm this, we measure both the lifetime and the coherence time of the single photons. As shown in Fig.$\,$3a, the photon lifetime at 4.3$\,$K (black square) is measured to be $T_1$$\,$$=$$\,$$162(5)$$\,$ps, which is shortened by a factor of 3.8 compared to the data at large detuning (red circle) due to Purcell effect. We use the Mach-Zehnder interferometer to measure the coherence time of the single photons. The integration time for each data point is 0.02$\,$s---a much longer timescale than those in the photon interference measurements. By fitting the Mach-Zehnder interference visibilities as a function of temporal delay (see Fig.$\,$3b), we extract the coherence time to be $T_2=294(8)$$\,$ps, and thus, $T_2/2T_1$$\,$$=$$0.91(5)$. This is in good agreement with the photon indistinguishability observed in the long timescale.

The emitted photons are further spectrally characterized by measuring their high-resolution spectra using a Fabry-Perot scanning cavity. The spectrum at $\pi$ pulse excitation is shown in Fig.$\,$3c where each data point is integrated for 0.1$\,$s, which is again a slow measurement compared to HOM interference. It can be slightly better fitted using a Voigt profile (red) than a pure Lorentzian (blue), with a homogeneous (Lorentzian) and inhomogeneous (Gaussian) linewidth of 1.01(4)$\,$GHz and 0.75(5)$\,$GHz, respectively. We expect that the Gaussian component could be due to inhomogeneous broadening caused by the spectral wandering around the center of the Lorentzian profile. Under this simple model, numerical analysis shows an average overlap of the photon wave packet of 0.90(2).

\begin{figure*}[tb]
    \centering
       \includegraphics[width=0.97\textwidth]{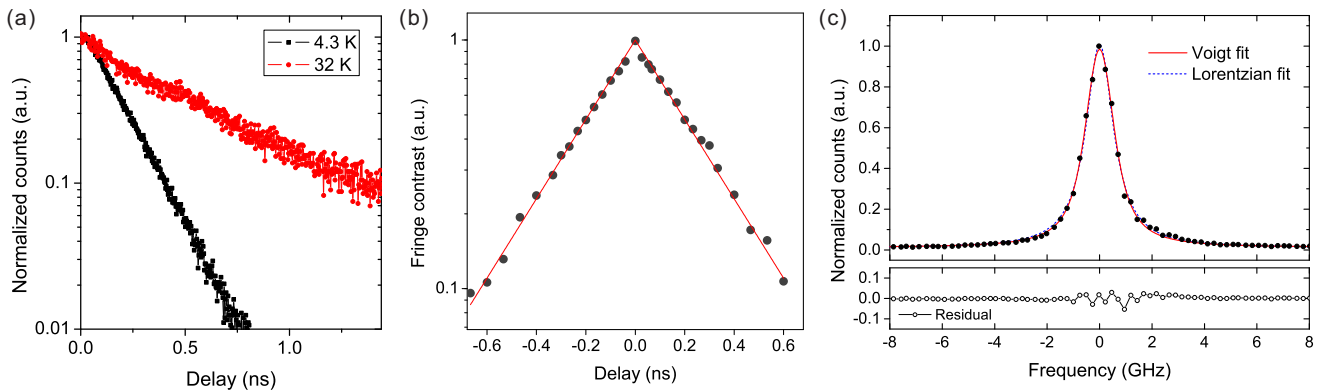}
\caption{Characterization of the single-photon source. (a) Time-resolved RF counts at QD-cavity resonance (4.3$\,$K) and at far detuning (32$\,$K). (b) Measurement of the coherence time of the single photons using a Mach-Zehnder interferometer, showing fringe contrast versus path-length difference. The fringe contrast was calculated from registered single photon counts of one of the interferometer outputs while scanning one of the arm lengths over about ten wavelengths. (c) A high-resolution RF spectrum when excited by a $\pi$ pulse, obtained using a home-built Fabry-P\'{e}rot scanning cavity with a finesse of 170, a linewidth of 220$\,$MHz (full width at half maximum), a free spectral range of 37.4$\,$GHz, and a total transmission rate of 61$\%$. The blue line was a fit using Lorentzian. The red line was a fit using Voigt profile with residual shown in the bottom panel. Fourier-transform this spectrum after deconvolution with the 220-MHz Fabry-P\'{e}rot instrumental resolution gives a coherence time of 291(8) ps, in good agreement with (b).}
\label{fig:3}
\end{figure*}

For the spectral measurement of our experiment, the Fabry-Perot scanning bandwidth is limited to $\sim\,$KHz for a reasonable signal to noise ratio, due to the total photon counts. If the scanning frequency can be faster than MHz, we would expect to see a narrowing of the spectrum and a transition from Voigt to purely Lorentzian.

The sub-$\mu{s}$ time scale of dephasing observed here is significantly longer than the previous results \cite{12.stephan,2013NatComm,2016NatPho}, meanwhile the spectral diffusion amplitude relative to the intrinsic lifetime-limited linewidth is smaller. Compared to the QDs grown by metal organic vapor phase epitaxy which is not carried out at ultra-high vacuum conditions \cite{12.stephan}, the results suggest our sample grown by molecular beam epitaxy may have smaller density of crystal impurities. Compared with the results that required stabilizing the electric environment around the QD with an additional laser \cite{2013NatComm} or electric gating \cite{2016NatPho}, we  conjecture that our QD-micropillar device in a high purity GaAs is exposed to a less noisy environment, such as an etched surface, interface or a crystal defect.

Controlled optical experiments are further performed to investigate the necessary condition for the generation of transform limited single photons from this QD. We use non-resonant excitation on the same QD and test two-photon interference using the same HOM setup. Firstly, we mix a small amount of 780$\,$nm laser---typically used for above band gap excitation---into the $s$-shell excitation laser. The power of the 780$\,$nm laser is on the order of tens of pW, which alone is too small to generate appreciable single photon counts. The non-resonant laser is filtered out in the output from the single photons using a long-pass filter. Figure$\,$S3 shows the dependence of photon indistinguishability as a function of the amount of mixed non-resonant laser. We observe a gradual decrease of indistinguishability with increasing amount of 780$\,$nm laser, which goes down to 54\% at 67$\,$pW.

Secondly, we test $p$-shell excitation on the QD where the laser is tuned to 881$\,$nm. With a high pump power of 120$\,$$\mu$W, the generated single photon rate is comparable to that under 21$\,$nW $\pi$-pulse resonant excitation (Fig.$\,$1c). Under this condition, we measure the intensity correlation and photon indistinguishability. As shown in Fig.$\,$S4, while the source still exhibits high single-photon purity: $g^2(0)$$\,$$=$$\,$$0.027(2)$, the interference visibility is measured to be 0.21(2), much lower than in $s$-shell excitation. The low visibility can be caused by high-power (120$\,\mu$W, as single-photon generation efficiencies increased asymptotically with pump power in incoherent excitation) laser excitation induced dephasing \cite{Bennett,EID}, and the time jitter from the non-radiative $p$- to $s$-shell incoherent relaxation \cite{Santori}. These tests indicate that $s$-shell pulsed resonant excitation appears essential for the generation of transform limited single photons on demand.

Finally, under strict resonant excitation, the photon indistinguishability are measured with the sample temperatures varying from 4.3$\,$K to 12$\,$K. It should be noted that the increase of temperature meanwhile brings the QD out of resonance with the micropillar cavity (see Fig.$\,$1b). Two-photon interference are tested with emission time separation of 13$\,$ns and 14.7$\,$$\mu$s. From the temperature-dependent result shown in Fig.$\,$S5, we observe an almost linear decrease of the indistinguishability with increasing temperature. This can be due to an combined effect of increased phonon-induced dephasing and reduced QD-cavity coupling \cite{phonon,PRB,Santori}. This test suggests that QD-microcavity devices with large Purcell enhancement at low temperature are desirable for emitting transform limited single photons.

In summary, we have tested time-dependent HOM interference between two single photons from a QD-micropillar, with a time separation up to 14.7$\,\mu$s, which not only elucidates the time dynamics of the dephasing process in single-photon generation, but also the quantitatively reveals the noise amplitude from the spectral diffusion. Streams of a thousand of mutually highly indistinguishable single photons are observed, lending them directly useful in boson sampling \cite{heyu} and optical quantum computing algorithms using the time-bin encoding protocol \cite{9.timebin}. Combining temporal correlation measurements and spectral analysis allows us to determine the single photons to be near transform limit ($>$$\,$90$\%$).

\textit{note}: After this work was completed, we became aware of a related paper on arXiv \cite{noteadd}.

\vspace{0.1cm}
\noindent \textit{Acknowledgement}: This work was supported by the National Natural Science Foundation of China, the Chinese Academy of Sciences, the National Fundamental Research Program, and the State of Bavaria.


\begin{thebibliography}{10}
\bibitem{details}
The wafer is grown via molecular beam epitaxy on a silicon doped GaAs wafer. On top of a 400 nm thick, intrinsic GaAs buffer layer, 25.5 (15) AlAs/GaAs mirror pairs which form the lower (upper) distributed Bragg reflector (DBR) are grown. The DBR stacks sandwich a $\lambda$-thick GaAs cavity with a single, low density layer of In(Ga)As QDs as active medium. The partial capping and annealing technique is applied to tune the QD emission wavelength to $\sim$900 nm. Underneath the QD layer, we implement a silicon $\delta$-doped layer with a dopant density of approximately $10^9$/cm$^2$. After growing is completed, micropillars with varying diameters ranging from 1 $\mu$m to 8 $\mu$m are defined via electron beam lithography and transferred into the sample by electron-cyclotron-resonance reactive ion-etching. The sample is then planarized by the polymer benzocyclobutene (BCB) and the etch mask is fully removed.
\end{thebibliography}
\end{document}